\documentclass[12pt]{article}

\usepackage{todonotes}

\usepackage{amsmath,amssymb,amsbsy,amsfonts,amsthm,latexsym,
                        amsopn,amstext,amsxtra,euscript,amscd,color}
\usepackage{color,xcolor}
\usepackage{latexsym}
\usepackage{cite}
\usepackage{amsfonts}
\usepackage{amsfonts,mathrsfs}
\usepackage{graphicx}
\usepackage{psfrag}
\usepackage{subfigure}
\usepackage{url}
\usepackage{stfloats}
\usepackage{amsmath}
\usepackage{algorithm}
\usepackage{algorithmic}
\usepackage{hyperref}

\hypersetup{colorlinks=true}

\newtheorem{theorem}{Theorem}
\newtheorem{lemma}{Lemma}

\newcommand{\quash}[1]{}

\begin{document}

\title{Arithmetic crosscorrelation of pseudorandom binary sequences of coprime periods}

\author{Zhixiong Chen$^{1}$, Zhihua Niu$^{1,2}$ and Arne Winterhof$^{3}$\\ \\
1. Key Laboratory of Applied Mathematics of Fujian Province University,\\
 Putian University, Putian, Fujian 351100, P. R. China\\
 ptczx@126.com\\
2. School of Computer Engineering and Science,\\
 Shanghai University, Shanghai 200444, P. R. China\\
zhniu@shu.edu.cn\\
3. Johann Radon Institute for Computational and Applied Mathematics,\\
Austrian Academy of Sciences, Altenberger Stra\ss e 69, A-4040 Linz, Austria \\
arne.winterhof@oeaw.ac.at
}

\maketitle

\begin{abstract}
The (classical) crosscorrelation is an important measure  of pseudorandomness of two binary sequences  for applications in communications.
The arithmetic crosscorrelation is another figure of merit introduced by  Goresky and Klapper generalizing Mandelbaum's arithmetic autocorrelation.

 First we observe
that the arithmetic crosscorrelation is constant for two binary sequences of coprime periods  which complements the analogous result for
the classical crosscorrelation.

 Then we prove upper bounds for the constant arithmetic crosscorrelation of two Legendre sequences of different periods and of two binary $m$-sequences of coprime periods, respectively.
\end{abstract}



\section{Introduction}

\indent

Sequences with good correlation properties are essential ingredients in a wide range
of applications including cryptography, CDMA systems and radar ranging \cite{GG}. A great deal of research
has gone into the design and generation of sequences and families of sequences
with good correlation properties. For example, for CDMA we need large families of
sequences with small pairwise correlations.

Let $N$ be the common (minimal) period of two  (periodic) binary  sequences
$$\mathcal{S}=(s_i)_{i\geq 0}\quad \mbox{and}\quad \mathcal{T}=(t_i)_{i\geq 0}$$
over the binary field $\mathbb{F}_2=\{0,1\}$.
The \emph{(classical) periodic crosscorrelation} of $\mathcal{S}$ and~$\mathcal{T}$ at lag $\tau$, denoted by $\mathcal{C}_{\mathcal{S,T}}(\tau)$, is defined by
$$
\mathcal{C}_{\mathcal{S,T}}(\tau)=\sum\limits_{0\leq i<N} (-1)^{s_i + t_{i+\tau}}, ~~~ 0\leq \tau<N.
$$
For $\mathcal{S}=\mathcal{T}$, it is called the \emph{(classical) periodic autocorrelation of $\mathcal{S}$  at $\tau$}, which is
denoted by
$$\mathcal{A}_{\mathcal{S}}(\tau)=\mathcal{C}_{\mathcal{S,S}}(\tau),\quad 0\le \tau<N.$$

A different notion of autocorrelation is the
  arithmetic autocorrelation introduced by Mandelbaum \cite{M1967} and later generalized to the arithmetic crosscorrelation by Goresky and Klapper~\cite{GK1997}. (Note that Mandelbaum did not use the term arithmetic autocorrelation.)
In the \emph{arithmetic crosscorrelation}, a sequence is added to a shift of another one
with carry, rather than bit by bit modulo $2$.
According to  \cite[Proposition 2]{GK1997} or
the discussions in \cite{HW2017,HMW2017}, we can demonstrate the computation (of the arithmetic crosscorrelation of  $\mathcal{S}$ and $\mathcal{T}$) as follows:

Write
$$
S(2)= \sum\limits_{0\leq i<N}s_i2^{i}, \quad  T^{(\tau)}(2)= \sum\limits_{0\leq i<N}t_{i+\tau}2^{i}.
$$
We compute $S(2)-T^{(\tau)}(2)$ in $\mathbb{Z}$. If $S(2)-T^{(\tau)}(2)\geq 0$, we   consider the unique binary expansion of $S(2)-T^{(\tau)}(2)$:
$$
S(2)-T^{(\tau)}(2)= \sum\limits_{0\leq i<N} w_i 2^i, ~~~ w_i\in\{0,1\}.
$$
If $S(2)-T^{(\tau)}(2)<0$, we  consider the binary expansion of $2^{N}-1+S(2)-T^{(\tau)}(2)\ge 0$:
$$
2^{N}-1+S(2)-T^{(\tau)}(2)= \sum\limits_{0\leq i<N} w_i 2^i, ~~~ w_i\in\{0,1\}.
$$
Then we compute the arithmetic crosscorrelation of $\mathcal{S}$ and $\mathcal{T}$ at $\tau$, denoted by $\mathcal{C}^{A}_{\mathcal{S,T}}(\tau)$:

\begin{equation}\label{CA}
\mathcal{C}^{A}_{\mathcal{S,T}}(\tau)=N_0-N_1=2N_0-N=N-2N_1,
\end{equation}
where for $j\in \{0,1\}$, $N_j$ is the number of $i=0,1,\ldots,N-1$ with $w_i=j$.
For $\mathcal{S}=\mathcal{T}$, this is the \emph{arithmetic  autocorrelation} of $\mathcal{S}$ at $\tau$,
denoted by
$$\mathcal{A}^{A}_{\mathcal{S}}(\tau)=\mathcal{C}^A_{\mathcal{S,S}}(\tau).$$
The reader is referred to the  research papers by
Goresky and Klapper \cite{GK1997,GK2008,GK2011} or  their monograph \cite{GK2012} for more background and results on arithmetic auto-/crosscorrelation.

 It is desirable that the absolute values of both the classical and the arithmetic cross-/autocorrelations are as small as possible  for $1\le \tau<N$.
However, sequences with small $\max\limits_{1\le\tau<N}|\mathcal{A}_{\cal S}(\tau)|$
may have large
$\max\limits_{1\le \tau<N}|\mathcal{A}_{\cal S}^A(\tau)|$
and  vice versa.

For example,
any $m$-sequence $\mathcal{S}$ produced by an $n$-order linear feedback shift register (LFSR),  that is of period $2^n-1$
satisfies
$$
\mathcal{A}_{\mathcal{S}}(\tau)=-1 \quad \mbox{and}\quad |\mathcal{A}^{A}_{\mathcal{S}}(\tau)|\leq 2^{n-1}-1,  \quad 1\leq \tau<2^n-1,
$$
see \cite[p.764]{H2011} and \cite{CNSW2021}.
 Numerical examples in \cite{CNSW2021} support the conjecture
$$\max_{1\le \tau< 2^n-1}|\mathcal{A}^{A}_{\mathcal{S}}(\tau)|=2^{n-1}-1.$$
In Section~\ref{sub-m-auto} below, we will give a bound on the  absolute value of the arithmetic autocorrelation function of $m$-sequences for small lags $\tau$, which is not considered in \cite{CNSW2021}.

Any $\ell$-sequence $\mathcal{S}$ produced by a feedback with carry shift register (FCSR) with the prime connection number $p$, that is of period $p-1$, can be defined by
$$
s_i=(a2^{-i} \mod p) \mod 2,
$$
for some $a\not\equiv 0 \pmod p$ and satisfies
 \begin{align*}
&\mathcal{A}_{\mathcal{S}}((p-1)/2)= p-1, \\
&\max_{\stackrel{1\le \tau< p-1}{\tau\not=(p-1)/2}}|\mathcal{A}_{\mathcal{S}}(\tau)|= EL_3(p),\\
&\mathcal{A}^{A}_{\mathcal{S}}(\tau)=0, \quad 1\leq \tau<p-1,
\end{align*}
where $EL_3(p)$ is the greatest even number less than $p/3$, see \cite[Theorem 5]{TQ2009} and \cite[Theorem 13.3.1]{GK2012}.

It is easy to see that
the classical crosscorrelation $\mathcal{C}_{\mathcal{S,T}}(\tau)$ is constant
if the periods of $\mathcal{S}$ and $\mathcal{T}$ are coprime.
Indeed, if $p$ is the period of $\mathcal{S}$ and $q$ is the period
of $\mathcal{T}$ with $\gcd(p,q)=1$, then we have  by the Chinese Remainder Theorem
$$
\mathcal{C}_{\mathcal{S,T}}(\tau)= \sum_{i=0}^{pq-1}(-1)^{s_i + t_{i+\tau}}=\sum_{i_1=0}^{p-1}(-1)^{s_{i_1}}\sum_{i_2=0}^{q-1}(-1)^{t_{{i_2}+\tau}},
$$
which is constant since the sum over $i_2$ is independent of $\tau$.
In particular, if both $\mathcal{S}$ and $\mathcal{T}$ are balanced with odd $pq$ (which is true for example for Legendre sequences  of period $p\equiv 3\pmod 4$ and $m$-sequences), we get $\left|\mathcal{C}_{\mathcal{S,T}}(\tau)\right|=1$ for $0\le \tau<pq$. If
one of the sequences $\mathcal{S}$ and $\mathcal{T}$ is balanced with even period, we get $\mathcal{C}_{\mathcal{S,T}}(\tau)=0$ for $0\le \tau<pq$.

In this work,  first we will prove  that
the arithmetic crosscorrelation
$\mathcal{C}^{A}_{\mathcal{S,T}}(\tau)$ is also constant under the same assumption  that the periods of $\mathcal{S}$ and $\mathcal{T}$  are coprime, see Section~\ref{sect-arith-cross}. Then we consider two Legendre sequences of different periods and two binary $m$-sequences of coprime periods and  derive upper bounds on their constant arithmetic crosscorrelation in Sections~\ref{sub-Legendre} and \ref{sub-m-cross}, respectively. Finally we  provide some numerical
 data in Section~\ref{sect-final}.

Occassionally, we will use negative indices for $N$-periodic sequences ${\cal S}=(s_i)_{i\ge 0}$ defined in the obvious way,
$$
s_{-i}=s_{N-i},\quad i=0,1,\ldots.
$$
We use the notation $f(n)=O(g(n))$ or $f(n)\ll g(n)$ if there is an absolute constant $c>0$ such that $|f(n)|\le cg(n)$.

\section{Arithmetic crosscorrelation}\label{sect-arith-cross}

In this section we prove the following result on the constant arithmetic crosscorrelation.

\begin{theorem}\label{Arith-Cross}
Let $\mathcal{S}=(s_i)_{i\geq 0}$ and $\mathcal{T}=(t_i)_{i\geq 0}$  be two sequences over $\mathbb{F}_2$ of periods $p>1$ and $q>1$, respectively.
If $p$ and $q$ are coprime, then the arithmetic crosscorrelation $\mathcal{C}^{A}_{\mathcal{S,T}}(\tau)$  has the same value for any $0\leq \tau<pq$.
\end{theorem}
Proof.
We consider the unique binary expansion of a non-negative integer $W<2^r-1$:
$$
W=w_0+w_12+\cdots+w_{r-1}2^{r-1},\quad w_i\in\{0,1\},~0\le i<r,
$$
and define the \emph{weight} of $W$, denoted by $wt(W)$, as
$$
wt(W)=\sum_{i=0}^{r-1}w_{i}.
$$
It is clear that
$$wt(2^kW)=wt(W)\quad \mbox{for any integer $k\geq 0$}.$$
Furthermore,
 for $k\ge 0$ let $W_k$ be defined by
$$W_k\equiv 2^k W\pmod {2^r-1},\quad 0\le W_k<2^r-1.$$
Then we have
\begin{eqnarray*}
W_k           &\equiv & w_02^k+w_12^{k+1}+\cdots+w_{r-1-k}2^{r-1}\\
           &&        +w_{r-k}+w_{r-k+1}2^{1}+\cdots+w_{r-1}2^{k-1} \pmod {2^{r}-1}.
\end{eqnarray*}
and thus
\begin{equation}\label{2kDelta}
wt(W_k)=wt(W),\quad k\ge 0.
\end{equation}

Since otherwise the result is trivial, we may assume that ${\cal T}$ is not constant.
The common minimal period of  $\mathcal{S}$ and $\mathcal{T}$ is
$N=pq$.
According to the definition of the arithmetic crosscorrelation,
we need to determine the weight of $S(2)-T^{(\tau)}(2)$ or $2^{pq}-1+S(2)-T^{(\tau)}(2)$ for $0\leq \tau<pq$.
We remark that  $$-(2^{pq}-1) < S(2)- T^{(\tau)}(2) < 2^{pq}-1,$$
since $0 \leq  S(2) \leq 2^{pq}-1$ and $0 < T^{(\tau)}(2) < 2^{pq}-1$.

Put
$$
\lambda_{k}=\sum_{m=0}^{p-1}(s_{m}-t_{m+kp})2^{m}, \quad k\geq 0.
$$
It is easy to see that
$$\lambda_{k+q}=\lambda_{k},\quad k\geq 0.$$

\textbf{Case 1.} We assume $S(2)-T^{(\tau)}(2)\geq 0$.\\
Take $x\in\{0,1,\ldots,q-1\}$ with $x\equiv \tau p^{-1}\pmod q$. We substitute
$$n=n_1+n_2p\quad \mbox{and}\quad n_3=x+n_2$$
to get
\begin{eqnarray*}
2^{xp}(S(2)-T^{(\tau)}(2))&\equiv&2^{xp} \sum_{n=0}^{pq-1} (s_n-t_{n+\tau})2^{n},\\
&\equiv& 2^{xp}\sum_{n_2=0}^{q-1}\left(\sum_{n_1=0}^{p-1}(s_{n_1}-t_{n_1+(x+n_2)p})2^{n_1}\right)2^{n_2p}.\\
&\equiv& \sum_{n_3=x}^{q-1+x} \left(\sum_{n_1=0}^{p-1} (s_{n_1}-t_{n_1+n_3p})2^{n_1}\right)2^{n_3p}\\
&\equiv& \sum_{n_3=x}^{q-1+x}\lambda_{n_3}2^{n_3p}\\
&\equiv & \sum_{n_3=0}^{q-1}\lambda_{n_3}2^{n_3p} \pmod {2^{pq}-1},
\end{eqnarray*}
since $\lambda_{k+q}=\lambda_k$.
We note that $\sum_{n_3=0}^{q-1}\lambda_{n_3}2^{n_3p}$ is 
independent of $\tau$ and we have a fixed number $\Omega$ with $0\leq \Omega < 2^{pq}-1$ such that
$$
\Omega \equiv \sum_{n_3=0}^{q-1}\lambda_{n_3}2^{n_3p} \pmod {2^{pq}-1}.
$$
Then by \eqref{2kDelta}, we derive for any $\tau$ with $S(2)-T^{(\tau)}(2)\geq 0$,
$$
wt(S(2)-T^{(\tau)}(2))=wt(2^{xp}(S(2)-T^{(\tau)}(2)))=wt(\Omega).
$$

\textbf{Case 2.} We assume $S(2)-T^{(\tau)}(2)< 0$.\\
We need to determine the weight of $2^{pq}-1+S(2)-T^{(\tau)}(2)$, which is a non-negative number smaller than $2^{pq}-1$.
Since for $x\in\{0,1,\ldots,q-1\}$ with $x\equiv \tau p^{-1}\pmod q$,
 $$
 2^{xp}(2^{pq}-1+S(2)-T^{(\tau)}(2))\equiv  2^{xp}(S(2)-T^{(\tau)}(2)) \pmod {2^{pq}-1},
 $$
we follow the proof in Case 1 to get
$$
2^{xp}(2^{pq}-1+S(2)-T^{(\tau)}(2))\equiv \sum_{n_3=0}^{q-1}\lambda_{n_3}2^{n_3p} \equiv \Omega \pmod {2^{pq}-1},
$$
from which we derive
$$
wt(2^{pq}-1+S(2)-T^{(\tau)}(2))=wt(2^{xp}(2^{pq}-1+S(2)-T^{(\tau)}(2)))=wt(\Omega)
$$
for any $\tau$ with $S(2)-T^{(\tau)}(2)< 0$.\\

Putting both cases together, we obtain by \eqref{CA}
\begin{eqnarray*}
\mathcal{C}^{A}_{\mathcal{S,T}}(\tau)&=&pq-\left\{
 \begin{array}{ll}
 2wt(S(2)-T^{(\tau)}(2)), & \mbox{if } S(2)-T^{(\tau)}(2)\geq 0,\\
 2wt(2^{pq}-1+S(2)-T^{(\tau)}(2)), & \mbox{otherwise,}
 \end{array}
 \right.\\
 &=&pq-2wt(\Omega).
\end{eqnarray*}
So $\mathcal{C}^{A}_{\mathcal{S,T}}(\tau)$ is constant which completes the proof. \qed

\section{Upper bounds for special  pairs of binary sequences}

In this section, we will prove upper bounds on the arithmetic crosscorrelation for two binary Legendre sequences of different periods and two binary $m$-sequences of coprime periods,  respectively. For results on their arithmetic autocorrelation, we refer the reader to \cite{CNSW2021,HW2017}.

\subsection{Arithmetic crosscorrelation of two binary Legendre sequences of different periods}
\label{sub-Legendre}

For a prime $p>2$ let $(\ell_n)$ be the \textit{Legendre sequence} defined by
	$$
		\ell_n=\begin{cases}
					1,&\text{if}~\left(\frac{n}{p}\right)=1;\\
					0,&\text{otherwise},
				\end{cases} \quad n\geq 0,
	$$
where $\left(\frac{.}{.}\right)$ is the Legendre symbol. Obviously, $(\ell_n)$ is $p$-periodic.

 We need a preliminary result on the pattern distribution of two Legendre sequences.
\begin{lemma}\label{lem-Legendre}
Let $p$ and $q$ be primes with $2<p<q$ and denote by $\mathcal{S}=(s_i)_{i\geq 0}$ and~$\mathcal{T}=(t_i)_{i\geq 0}$
the Legendre sequences of periods $p$ and $q$, respectively.

For an integer $k\ge 0$ and any pattern $\underline{e}\in \{0,1\}^{2k+2}$, the number $\Sigma_k$ of
$i=0,1,\ldots,pq-1$ with
$$(s_{i-k},s_{i-k+1},\ldots,s_i,t_{i-k},t_{i-k+1},\ldots,t_i)=\underline{e}$$
satisfies
$$\Sigma_k=\frac{pq}{2^{2k+2}}+O\left(2^{-k}kp^{1/2}q\right)\quad \mbox{for} \quad k\le (0.5 \log p-\log\log p)/\log 2.$$
\end{lemma}
Proof. 
Write $\underline{e}=(e_0,\ldots,e_{2k+1})$ and
put
$$\delta_{i,j,p}=1-(-1)^{e_{k-j}}\left(\frac{i-j}{p}\right)\quad \mbox{and}\quad \delta_{i,j,q}=1-(-1)^{e_{2k+1-j}}\left(\frac{i-j}{q}\right).$$
Note that for $i$ and $j\le k$ with $\gcd(i-j,pq)=1$ we have
$$\delta_{i,j,p}\delta_{i,j,q}
=\left\{\begin{array}{cl} 4, &\mbox{if }(s_{i-j},t_{i-j})=(e_{k-j},e_{2k+1-j}),\\
0, & \mbox{otherwise,}\end{array}\right.$$
and thus
$$\Sigma_k=\frac{1}{2^{2k+2}}\sum_{i=0}^{pq-1} \prod_{j=0}^k \delta_{i,j,p}\delta_{i,j,q}+O(kq),$$
where the $O$-term comes from those $i$ with $\gcd(i-j,pq)>1$ for some $j=0,1,\ldots,k$.
Expanding the product we get
$$\Sigma_k=\frac{1}{2^{2k+2}}\sum_{U,V\subseteq\{0,1,\ldots,k\}}\sum_{i=0}^{pq-1}\prod_{j\in U} (-1)^{e_{k-j}}\left(\frac{i-j}{p}\right)\prod_{j\in V}(-1)^{e_{2k+1-j}}\left(\frac{i-j}{q}\right)+O(kq).$$

The contribution to $\Sigma_k$ of $U=V=\emptyset$ is trivially
$$\frac{pq}{2^{2k+2}}.$$
The contribution for $U=\emptyset$ and the $(2^{k+1}-1)$ sets  $V\not=\emptyset$
is bounded by
$$
\frac{1}{2^{2k+2}}\cdot (2^{k+1}-1)\cdot p\max_{V\not=\emptyset}\left|\sum_{i=0}^{q-1}\left(\frac{\prod\limits_{j\in V}(i-j)}{q}\right)\right|=O\left(2^{-k}kpq^{1/2}\right)
$$
by the Weil bound,  see for example \cite[Theorem~5.41]{LN1997}.
Analogously, the contribution to $\Sigma_k$ for $V=\emptyset$ and the $(2^{k+1}-1)$ sets $U\not=\emptyset$ is
$$O\left(2^{-k}kp^{1/2}q\right).$$
The contribution of the $2^{2k+2}-2^{k+2}+1$
remaining $(U,V)$ with $U\not=\emptyset$ and $V\not=\emptyset$ is bounded by
\begin{eqnarray*}
&&\frac{1}{2^{2k+2}}\cdot (2^{2k+2}-2^{k+2}+1)\cdot \max_{U,V\not=\emptyset}\left|\sum_{i=0}^{pq-1}\prod_{j\in U}\left(\frac{i-j}{p}\right)\prod_{j\in V}\left(\frac{i-j}{q}\right)\right|\\
&\le&\max_{U,V\not=\emptyset}\left|\sum_{i_1=0}^{p-1}\left(\frac{\prod\limits_{j\in U}(i_1-j)}{p}\right)\sum_{i_2=0}^{q-1}\left(\frac{\prod\limits_{j\in V}(i_2-j)}{q}\right)\right|\\
&=&O(k^2 (pq)^{1/2})
\end{eqnarray*}
by the Chinese Remainder Theorem and the Weil bound.
Collecting everything and verifying that
$$\max\{kq,2^{-k}kp^{1/2}q,k^2(pq)^{1/2}\}=2^{-k}kp^{1/2}q\quad \mbox{for}\quad k\le (0.5 \log p-\log\log p)/\log 2$$
we get the result. \qed

\begin{theorem} \label{thm-Legendre-bound}
Let $\mathcal{S}=(s_i)_{i\geq 0}$ and $\mathcal{T}=(t_i)_{i\geq 0}$ be two Legendre sequences over $\mathbb{F}_2$ of prime periods $p>2$ and $q>2$, respectively.
If $p<q$, then the arithmetic crosscorrelation $\mathcal{C}^{A}_{\mathcal{S,T}}(\tau)$ satisfies
$$
\mathcal{C}^{A}_{\mathcal{S,T}}(\tau)\ll p^{1/2}q (\log p)^{2},\quad 0\leq \tau<pq.$$
\end{theorem}
Proof. By Theorem \ref{Arith-Cross} we may assume $\tau=0$.
Without loss of generality we may assume  $S(2)\ge T^{(0)}(2)$. We write
$$S(2)-T^{(0)}(2)=\sum_{i=0}^{pq-1}(s_i-t_i)2^i=\sum_{i=0}^{pq-1}w_i2^i \quad\mbox{with}\quad w_i\in \{0,1\}.$$
Note that the $w_i$ are unique and we have to estimate the number of $i=0,1,\ldots,pq-1$ with $w_i=1$.

Assume that for some $n\ge k\ge 1$ and $a\in \{0,1\}$
	\begin{align*}
			& (s_{n-k},t_{n-k})=(a,1-a), \\
			&s_{n-k+j}=t_{n-k+j},\quad j=1,\ldots,k-1,\\
			&(s_n,t_n)\in \{0,1\}^2.
	\end{align*}

For $a=1$ we have
$$2^{n-k+1}>\sum_{i=0}^{n-k} (s_i-t_i)2^i\ge 2^{n-k}-\sum_{i=0}^{n-k-1}2^i>0$$
and thus $w_{n-k+1}$ depends only on $(s_{n-k+1},t_{n-k+1})$. Obviously, we have
$$w_{n-k+j}=s_{n-k+j}-t_{n-k+j}=0\quad\mbox{for}\quad j=1,\ldots,k-1$$
and $w_n=1$ if and only if $s_n\not=t_n$.

For $a=0$ we have
$$0<2^{n-k+1}+\sum_{i=0}^{n-k}(s_i-t_i)2^i<2^{n-k+1}$$
and thus
$$w_{n-k+j}=1+s_{n-k+j}-t_{n-k+j}=1\quad \mbox{for}\quad  j=1,\ldots,k-1$$
and $w_n=1$ if and only if $s_n=t_n$.

Altogether there are $2^{k+1}$ different patterns $\underline{e}\in \{0,1\}^{2k+2}$ such that
\begin{equation}\label{=e}
(s_{n-k},s_{n-k+1},\ldots,s_n,t_{n-k},t_{n-k+1},\ldots,t_n)=\underline{e}
\end{equation}
implies $w_n=1$.
For each of these  $2^{k+1}$ patterns $\underline{e}$ there are
$$\frac{pq}{2^{2k+2}}+O\left(2^{-k}kp^{1/2}q\right)$$ different $n=k,k+1,\ldots,k+pq-1$ satisfying \eqref{=e}  by Lemma \ref{lem-Legendre}.
Hence, for each $k=1,2,\ldots$ there are at least
$$ 2^{k+1}\frac{pq}{2^{2k+2}}+O\left(\frac{2^{k+1}}{2^k}kp^{1/2}q\right)=\frac{pq}{2^{k+1}}+O\left(kp^{1/2}q\right)$$
different $n$  in each fixed interval of length $pq$ with $w_n=1$.
Choose
$$M=\left\lfloor \frac{\log p}{2\log 2}-\frac{2\log\log p}{\log 2}\right\rfloor\le \log p.$$
Summing up, the number $N_1$ of $n=0,1,\ldots,pq-1$ with $w_n=1$ is at least
\begin{eqnarray*}
N_1&\ge&  \sum_{k=1}^M\frac{pq}{2^{k+1}}+O\left(\sum_{k=1}^M kp^{1/2}q\right)\\
&=& \frac{pq}{4}\sum_{k=0}^{M-1}2^{-k} +O\left(M^2p^{1/2}q \right)\\
&=&\frac{pq}{2}\left(1-\left(\frac{1}{2}\right)^M\right)+O(p^{1/2}q(\log p)^2)\\
&=&\frac{pq}{2}+O(p^{1/2}q(\log p)^{2}),
\end{eqnarray*}
 where in the last step we used
$$p^{-1/2}(\log p)^2\le 2^{-M}\le 2 p^{-1/2}(\log p)^2$$
by the choice of $M$.

Similar, we can show that the number $N_0$ of $n$ with $w_n=0$ is at least
$$N_0\ge \frac{pq}{2}+O(p^{1/2}q(\log p)^{2}).$$
Since
$$N_0=pq-N_1\le \frac{pq}{2}+O(p^{1/2}q(\log p)^{2})$$ we get the result by \eqref{CA}.  \qed\\

 Put $N=pq$. In the important case that $p$ and $q$ are of the same order of magnitude the bound is of order of magnitude
$N^{3/4}(\log N)^{2}$.

\subsection{Arithmetic crosscorrelation of two binary $m$-sequences of coprime periods}
\label{sub-m-cross}

 First note that
$$d=\gcd(n_1,n_2)=1\quad \mbox{if and only if}\quad t=\gcd(2^{n_1}-1,2^{n_2}-1)=1.$$
This can be easily verified. On the one hand if $d>1$, then
$$2^{n_i}-1=\left(2^d-1\right)\left(1+2^d+\cdots+2^{(n_i/d -1)d}\right), \quad i=1,2,$$
and $2^d-1$ is a nontrivial divisor of $t$.
On the other hand if $t>1$, then there is a prime divisor $p>2$ of $t$ and the order of $2$ modulo $p$ divides $d$.

Let $g_n$ be a primitive element of the finite field $\mathbb{F}_{2^n}$. Then the sequence of the form\footnote{In fact, any $m$-sequence can be defined as $\overline{s}_i={\rm Tr}_n(ag_n^i),~ i=0,1,\ldots$, for some $0\not= a\in\mathbb{F}_{2^n}$, which is a shift of $s_i={\rm Tr}_n(g_n^i)$.}
$$s_i={\rm Tr}_n(g_n^i),\quad i=0,1,\ldots$$
is an $m$-sequence of period $2^n-1$, where
$${\rm Tr}_n(c)=c+c^2+\cdots+c^{2^{n-1}}, \quad c\in \mathbb{F}_{2^n},$$
denotes the (absolute) trace of $\mathbb{F}_{2^n}$.

We need a result on the pattern distribution of two $m$-sequences.
\begin{lemma}\label{lem-m-sequence}
Let $n_1<n_2$ be two coprime positive integers and denote by $\mathcal{S}=(s_i)_{i\geq 0}$ and $\mathcal{T}=(t_i)_{i\geq 0}$ two $m$-sequences
of periods $2^{n_1}-1$ and $2^{n_2}-1$, respectively.

For an integer $k\ge 0$ and any pattern $\underline{e}\in \{0,1\}^{2k+2}$, the number $\Sigma_k$ of
$i=0,1,\ldots,(2^{n_1}-1)(2^{n_2}-1)-1$ with
$$(s_{i-k},s_{i-k+1},\ldots,s_i,t_{i-k},t_{i-k+1},\ldots,t_i)=\underline{e}$$
satisfies
$$
\left|\Sigma_k-\frac{(2^{n_1}-1)(2^{n_2}-1)}{2^{2k+2}}\right|\le 2^{n_1-k-1} + 2^{n_2-k-1} +1\quad \mbox{for}\quad k<n_1.
$$
\end{lemma}
Proof.
Let $$s_i={\rm Tr}_{n_1}(g_{n_1}^i), \quad i=0,1,\ldots$$
for a primitive element $g_{n_1}$ of $\mathbb{F}_{2^{n_1}}$ and
$$t_i={\rm Tr}_{n_2}(g_{n_2}^i), \quad i=0,1,\ldots$$ for a primitive element $g_{n_2}$ of $\mathbb{F}_{2^{n_2}}$, respectively.

Put
$$N=(2^{n_1}-1)(2^{n_2}-1)$$
and
$$\delta_{i,j,n_1}=1+(-1)^{e_{k-j}}\psi_{n_1}(g_{n_1}^{i-j})\quad \mbox{and}\quad \delta_{i,j,n_2}=1+(-1)^{e_{2k+1-j}}\psi_{n_2}(g_{n_2}^{i-j}),$$
where
$$\psi_n(c)=(-1)^{{\rm Tr}_n(c)},\quad c\in \mathbb{F}_{2^n},$$
is the additive canonical character of $\mathbb{F}_{2^n}$.
Note that
$$\delta_{i,j,n_1}\delta_{i,j,n_2}=\left\{\begin{array}{cl} 4, &\mbox{if }(s_{i-j},t_{i-j})=(e_{k-j},e_{2k+1-j}),\\
0, & \mbox{otherwise,}\end{array}\right.$$
and thus we have
$$\Sigma_k=\frac{1}{2^{2k+2}}\sum_{i=0}^{N-1} \prod_{j=0}^k \delta_{i,j,n_1}\delta_{i,j,n_2}.$$
Expanding the product we get
$$\Sigma_k=\frac{1}{2^{2k+2}}\sum_{U,V\subseteq\{0,1,\ldots,k\}}\sum_{i=0}^{N-1}\prod_{j\in U} (-1)^{e_{k-j}}\psi_{n_1}(g_{n_1}^{i-j})\prod_{j\in V}(-1)^{e_{2k+1-j}}\psi_{n_2}(g_{n_2}^{i-j}).$$
The contribution of $U=V=\emptyset$ is
$$\frac{N}{2^{2k+2}}.$$
The contribution of $U=\emptyset$ and the $(2^{k+1}-1)$ sets $V\not=\emptyset$ is bounded by
$$2^{n_1-k-1}\max_{V\not=\emptyset}\left|\sum_{i=0}^{2^{n_2}-2}\psi_{n_2}\left(\sum_{j\in V}g_{n_2}^{-j}g_{n_2}^i\right)\right|.$$
Since $k<n_1<n_2$
and the minimal polynomial of $g_{n_2}^{-1}$ is of degree $n_2$, we have
$$\sum_{j\in V}g_{n_2}^{-j}\ne 0$$
and the sum over $i$ equals $-1$. So the contribution of $U=\emptyset$ and $V\not=\emptyset$ is at most
$$2^{n_1-k-1}.$$
Similarly we see that the contribution of $V=\emptyset$ and $U\not=\emptyset$ is
$$2^{n_2-k-1}.$$
In the remaining case the contribution of the $2^{2k+2}-2^{k+2}+1$ pairs of sets $(U,V)$ with $U\not=\emptyset$ and $V\not=\emptyset$ is at most $1$.

Putting everything together, we complete the proof. \qed

\begin{theorem}
Let $\mathcal{S}=(s_i)_{i\geq 0}$ and $\mathcal{T}=(t_i)_{i\geq 0}$ be two binary $m$-sequences over $\mathbb{F}_2$ of periods $2^{n_1}-1$ and $2^{n_2}-1$, respectively. If $n_1<n_2$  and $\gcd(n_1,n_2)=1$, then the arithmetic crosscorrelation $\mathcal{C}^{A}_{\mathcal{S,T}}(\tau)$ satisfies
$$
\mathcal{C}^{A}_{\mathcal{S,T}}(\tau)\ll n_12^{n_2},\quad
0\leq \tau< (2^{n_1}-1)(2^{n_2}-1).
$$
\end{theorem}
Proof. As in the proof of Theorem \ref{thm-Legendre-bound}
and using Lemma~\ref{lem-m-sequence}
we get
\begin{eqnarray*}N_i&\ge& \frac{(2^{n_1}-1)(2^{n_2}-1)}{2}\sum_{k=1}^{n_1-1}2^{-k} -(n_1-1)(2^{n_1}+2^{n_2}+1)\\
&= &\frac{(2^{n_1}-1)(2^{n_2}-1)}{2}+O(n_12^{n_2})
\end{eqnarray*}
for $i=0,1$.
Hence,
$$|N_0-N_1|\ll n_12^{n_2}$$
and the result follows.
\qed\\

The theorem above indicates that the arithmetic crosscorrelation of two $m$-sequences is quite small.
In particular, if $n_1$ and $n_2$ are close and $N=(2^{n_1}-1)(2^{n_2}-1)$, then the bound is of order of magnitude
$N^{1/2}\log N$ which is in good correspondence with the expected value of the absolute value of the arithmetic crosscorrelation of two random sequences of period $N$, see \cite[Theorem~8.3.6]{GK2012}.

However, the arithmetic autocorrelation of $m$-sequences is quite large.
Numerical data indicates that
its maximum value
is the greatest number less than half of the period \cite{CNSW2021}. To improve
the results in \cite{CNSW2021}, in the following
subsection, we will estimate the arithmetic autocorrelation function of $m$-sequences for small lags $\tau$.

\subsection{Arithmetic autocorrelation function of $m$-sequences for small lags $\tau$}\label{sub-m-auto}

By \cite[Proposition~2.1]{HW2017} we have for any $N$-periodic sequence
\begin{equation}\label{symm}{\cal A}_{\cal S}^A(\tau)=-{\cal A}_{\cal S}^A(N-\tau),\quad \tau=1,2,\ldots,N-1.
\end{equation}

Now we state a result on the pattern distribution of an $m$-sequence of period~$2^n-1$.

\begin{lemma}\label{lem-m-auto}
Let $\mathcal{S}=(s_i)_{i\geq 0}$ be an $m$-sequence of period $2^{n}-1$. Choose $\tau$ with $1\leq \tau<n$.

(1) For $k\ge 0$ and  any non-zero pattern $\underline{e}\in\{0,1\}^{2k+2}\setminus\{(0,0,\ldots,0)\}$, the number $\Sigma_k$ of
$i=0,1,\ldots,2^n-2$ with
$$(s_{i-k},s_{i-k+1},\ldots,s_i,s_{i-k+\tau},s_{i-k+\tau},\ldots,s_{i+\tau})=\underline{e}$$
is
$$\Sigma_k=2^{n-2k-2},\quad  k\le \min\{\tau,n-\tau\}-1.$$

(2) For $k\ge \tau$ and non-zero pattern $\underline{e}\in \{0,1\}^{k+\tau+1}\setminus\{(0,0,\ldots,0)\}$,  the number~$\sigma_k$ of $i=0,1,\ldots,2^n-2$ with
$$(s_{i-k},s_{i-k+1},\ldots,s_{i+\tau})=\underline{e}$$
is
$$\sigma_k=2^{n-k-\tau-1},\quad \tau\le k\le n-\tau-1.$$
\end{lemma}
Proof. For $\ell=1,2,\ldots,n$, we see that each non-zero pattern of length $\ell$ occurs as $(s_i,s_{i+1},\ldots,s_{i+\ell-1})$ for exactly $2^{n-\ell}$ different $i$ with $0\le i<2^n-1$, see for example \cite[Proposition~5.2]{GG}.
We choose $\ell=k+\tau+1\le n$ and (2) follows.

For (1) note that the choice of $(s_{i+1},s_{i+2},\ldots,s_{i-k+\tau-1})$ is free. Hence, we derive
$2^{n-\ell+\tau-k-1}=2^{n-2k-2}$ different $i$ with the desired pattern property.  \qed

\begin{theorem}
Let $\mathcal{S}=(s_i)_{i\geq 0}$ be an $m$-sequence of period $2^{n}-1$.
We have
$$\left|{\cal A}_{\cal S}^A(\tau)\right|\le 2^{\min\{n-1,\tau+1,2^n-\tau\}}-1,\quad 1\leq \tau<2^n-1.$$
\end{theorem}
Proof. The bound
$$\left|{\cal A}_{\cal S}^A(\tau)\right|\le 2^{n-1}-1$$ follows from \cite{CNSW2021}. By \eqref{symm} it remains to show
$$\left|{\cal A}_{\cal S}^A(\tau)\right|\le 2^{\tau+1}-1\quad
\mbox{for}\quad \tau\le n-3.$$
As before, let $N_j$ be the number of digits equal to $j\in\{0,1\}$ in the binary expansion of the integer $S(2)-S^{(\tau)}(2)$.
As in the above proofs, we get by Lemma \ref{lem-m-auto}
$$N_1\ge \sum_{k=1}^{\min\{\tau,n-\tau\}-1}2^{k+1}\Sigma_k+\sum_{k=\min\{\tau,n-\tau\}}^{n-\tau-1}2^\tau\sigma_k=\sum_{k=1}^{n-\tau-1}2^{n-k-1}=2^{n-1}-2^{\tau}$$
as well as $N_0\ge 2^{n-1}-2^{\tau}$ and thus
$$N_j\le 2^{n-1}+2^{\tau}-1,\quad j=0,1.$$
Hence, $|N_0-N_1|\le 2^{\tau+1}-1$ and \eqref{CA} finishes the proof.
\hfill \qed

 \section{Final remarks}\label{sect-final}

Below we list some numerical data for the arithmetic crosscorrelations of two Legendre sequences of coprime periods in Table \ref{table-Legendre}
and two binary $m$-sequences of coprime periods in Table \ref{table-m-sequence}.

\begin{table}[h]
\centering
\begin{tabular}{ccc}
\hline\noalign{\smallskip}
$p$   &   $q$          & $\mathcal{C}^{A}_{\mathcal{S,T}}(\tau)$  \\
 \noalign{\smallskip} \hline \noalign{\smallskip}
$7$ & $11$             &  $-1$               \\
$7$ & $13$             &  $5$                \\
$7$ & $17$             &  $-13$              \\
$7$ & $23$             &  $1$                \\
$11$ & $13$            &  $-3$               \\
$11$ & $19$            &  $-5$                \\
$13$ & $17$            &  $-7$              \\
$17$ & $29$            &  $9$                \\
\hline
\end{tabular}
\caption{Arithmetic crosscorrelation of two Legendre sequences $\mathcal{S}$  and $\mathcal{T}$ of periods $p$ and $q$}\label{table-Legendre}
\end{table}

Let us denote by $M_\mathcal{S}(X)\in \mathbb{F}_2[X]$ the minimal polynomial of $\mathcal{S}$, see \cite{CDR1998} for details.

\begin{table}[h]
\centering
\begin{tabular}{ccc}
\hline\noalign{\smallskip}
 $M_\mathcal{S}(X)$    &  $M_\mathcal{T}(X)$   &  $\mathcal{C}^{A}_{\mathcal{S,T}}(\tau)$  \\
 \noalign{\smallskip} \hline \noalign{\smallskip}
$X^3+X^2+1$ & $X^4+X^3+1$                    &  $-1$     \\
$X^3+X^2+1$ & $X^5+X^3+1$                    &   $1$     \\
$X^3+X^2+1$ & $X^7+X^6+1$                    &  $-3$     \\
$X^3+X^2+1$ & $X^8+X^6+X^5+X^4+1$            &  $7$      \\
$X^5+X^3+1$ & $X^8+X^6+X^5+X^4+1$            &  $-1$     \\
$X^6+X^5+1$ & $X^7+X^6+1$                    &  $-1$     \\
$X^7+X^6+1$ & $X^8+X^6+X^5+X^4+1$            &  $-3$     \\
\hline
\end{tabular}
\caption{Arithmetic crosscorrelation of two binary $m$-sequences $\mathcal{S}$ and~$\mathcal{T}$
with minimal polynomials $M_\mathcal{S}(X)$ and $M_\mathcal{T}(X)$}\label{table-m-sequence}
\end{table}

These tables indicate that the size of $\mathcal{C}^{A}_{\mathcal{S,T}}(\tau)$ can be quite different for different periods of similar size.

For two binary sequences $\mathcal{S}$ of period $p$ and $\mathcal{T}$ of period $q$  with $\gcd(p,q)>1$, their classical and arithmetic crosscorrelations are both not constant. For example,
if $\mathcal{S}$ is an $m$-sequence with minimal polynomial $X^4+X^3+1$ and $\mathcal{T}$  is an $m$-sequence with minimal polynomial $X^4+X+1$, we compute that
$$
\mathcal{C}_{\mathcal{S,T}}(\tau)\in\{-1,-5,3,7\}, \quad \mathcal{C}^{A}_{\mathcal{S,T}}(\tau)\in\{-3,-7,-9,1,3,5\}.
$$
So it would be interesting to  find an upper bound on $\mathcal{C}^{A}_{\mathcal{S,T}}(\tau)$
when the period of $\mathcal{S}$ is not coprime to that of $\mathcal{T}$.
For $m$-sequences the method of this paper still provides non-trivial results for very small and very large lags $\tau$
but fails for most $\tau$.

\section*{Acknowledgments}

  Z. Chen was partially supported by the National Natural Science
Foundation of China under grant No.~61772292, and by the Provincial Natural Science
Foundation of Fujian, China under grant No.~2020J01905.

Z. Niu  was partially supported by the National Key Research and Development Program of China under grant No.~2018YFB0704400, and by the
open program of Key Laboratory of Applied Mathematics of Fujian Province University (Putian University) under grant No.~SX202102.

\end{document}